# Kinetics of Ion Transport in Ionic Liquids: Two Dynamical Diffusion States

Guang Feng[1]*, Ming Chen[1], Sheng Bi[1], Zachary A.H. Goodwin[2,3], Eugene B. Postnikov[4], Michael Urbakh[5]*, and Alexei A. Kornyshev[3]*

[1]State Key Laboratory of Coal Combustion, School of Energy and Power Engineering, Huazhong University of Science and Technology (HUST), Room 321, Power Building, 1037 Luoyu Road, Wuhan, Hubei 430074 China

[2]Department of Physics, Faculty of Natural Sciences, Imperial College London, SW7 2AZ London, United Kingdom

[3]Department of Chemistry, Faculty of Natural Sciences, Imperial College London, SW7 2AZ London, United Kingdom

[4]Theoretical Physics Department, Kursk State University, Radishcheva Str.33, Kursk 305000, Russia

[5]School of Chemistry, The Sackler Faculty of Science, Tel Aviv University, Ramat Aviv, 69978 Tel Aviv, Israel

*Correspondence to: gfeng@hust.edu.cn, urbakh@post.tau.ac.il and a.kornyshev@imperial.ac.uk

**Abstract** (176 words-from 250 allowed)

Using classical molecular dynamics simulations, we investigate the mobility of ions in [Bmim][TFSI], a typical room temperature ionic liquid. Analyzing the trajectories of individual cations and anions, we estimate the time that ions spend in bound, 'clustered' states, and when the ions move quasi-freely. Using this information, we evaluate the average portion of 'free' ions that dominate conductivity. The amount of thus defined free ions comprises 15-25%, monotonically increasing with temperature in the range of 300-600 K, with the rest of the ions being temporarily bound, moving rather in local potentials. The conductivities as a function of temperature, calculated from electric current autocorrelation functions, reproduce reported experimental data well. Interestingly, for 'free' ions the Nernst-Einstein relationship between the mobility and diffusion coefficient holds fairly well. In analogy with electronic semiconductors, one can speak about an '*ionic semiconductor*' model for ionic liquids with *valence* (or 'excitonic') and *conduction* band states for ions, separated by an energy gap. The obtained band gap for the ionic liquid is, however, very small, ~0.026 eV, allowing for easy interchanges between the two dynamic states.

**Keywords:** ionic liquid | ion dynamics | ionic semiconductor | modified Nernst-Einstein relationship | molecular dynamics modeling

The rediscovery of **r**oom **t**emperature **i**onic **l**iquids (RTILs) – solvent free electrolytes not freezing at room temperature – was a revolution in chemistry (1). RTILs have been appraised first not as electrolytes but as almost universal solvents for synthesis and catalysis (1), dissolving depending on their composition, practically any substance of interest. In the past we could 'count' solvents, but now with RTILs there are practically an unlimited number of them. Easily mixable with each other, 'cocktails' of RTILs offered additional variety of 'designer solvents' (2, 3). However, RTILs are of great interest not only to synthetic chemists, but also to electrochemists and engineers (4-6). These electrolytes sustain higher applied voltages than aqueous electrolytes, with their ions not taking part in electrochemical reactions at electrodes when the solutes dissolved in them already do, which can speed up electrocatalytic reactions of the solutes. Together with nonvolatility and their sustainability to higher temperatures, RTILs were particularly interesting as electrolytes for electroctatalysis as well as in various energy-related devices, from supercapacitors to fuel cells and even batteries (4-6). For a physicist, RTILs comprise a unique kind of matter: room temperature dense ionic plasmas with strong Coulomb correlations.

Understanding RTILs at electrified interfaces, necessary for any electrochemical applications, required reconsideration of the theory of the electrical double layer (7). The first major attempt to do this, with discussions of avenues of further development, was presented in Ref. (8). Based on the simplest possible mean-field theory, it was demonstrated that RTILs could have dramatically different voltage dependence of the double layer capacitance from that prescribed by Gouy-Chapman theory of dilute electrolytes (9, 10). 'Bell' and 'camel' shaped capacitance-potential curves, instead of Gouy-Chapman U-shape, were obtained because of a restriction on the local increase of ionic concentration in response to electrode charging. The key parameter there (and practically the only one) was the ratio of average concentration of ions in the bulk to the maximum possible local concentration of ions, $\gamma$, referred to as *compacity* (7). The smaller $\gamma$, the stronger the increase of the concentration of counterions in the double layer could be.

At $\gamma =1/3$ the theory predicted a crossover from the bell to a camel shape, in which the capacitance first grows and then decreases as the concentration of ions saturates at the level of maximal possible concentration.

Ref. (8) did not dwell much on the specific value of $\gamma$, but rather considered it to be a parameter of the liquid, which could be related with the free volume due to voids between ions in such complex fluids. Skipping the history of the following development of the theory of the double layer in RTILs (7), we consider the latest views on the meaning of $\gamma$. These were influenced by two factors. Many estimates have shown that the volume portion of natural voids in RTILs is pretty small, just a few percentage (11, 12). Had the value of compacity been due to the physical voids between ions, $\gamma$ would not be less than 0.95. The next and even stronger impetus for reconsidering the meaning of $\gamma$ came from a 'disruptive' publication (13). In that work, using **s**urface **f**orce **a**pparatus (SFA) the Israelachvili group measured forces between a spontaneously charged mica surface and gold with an applied voltage separated by RTILs. A striking result was reported: at large distances between the surfaces the force seemed to obey the law of an exponentially decaying repulsion of the two overlapping electrical double layers, but with an extraordinary long decay length earlier observed only for extremely diluted electrolytes. A conjecture was raised to rationalize the data (14), that *pure RTILs are effectively dilute electrolytes*, in which most of the ions are bound into neutral clusters (ion pairs or larger aggregates), with only a minute amount of ions participating in screening. Once adopting this picture, the rest is straightforward:

(i) Large clusters of ions are majorly neutral, and have internal polarizability that provides dielectric screening to the 'free' ions;

(ii) As, hypothetically, only free ions can participate in screening and contribute to the formation of electrical double layers, a very small number of such ions would give rise to extra-long screening length.

Initially this conjecture met resistance. Later Perkin et al. obtained similar results measuring structural forces between mica surfaces in contact with simple electrolytes and ionic liquid solutions of different concentrations (15). Israelachvili, Perkin, Atkin, Rutland, and co-workers have then published a mini-review about this phenomenon (16). Perkin's group has put these findings in a broader picture of nonmonotonic dependence of the screening length on the electrolyte concentration in concentrated electrolytes (17). Indeed, in dilute ionic solutions most ions are free and contribute to screening, thus the increase of their concentration makes the Debye length shorter, but with further increase of concentration more ions get bound into pairs and larger clusters and thus stop contributing to Debye screening, the latter increasing again. Such ion pairing/clustering concepts in electrolytic solutions were developed from old to modern times (18-20).

But is it true that most of the ions are indeed clustered in pure RTILs? Various estimates (19, 21, 22) suggested that the expected degree of clustering is not enough to explain the observations of Gebbie et al. in such a simple way. Thus, the fact that both Perkin's and Atkin & Rutland's groups have approved the ultra-long-range forces between charged surfaces, still may not exclude that the explanation of these observations lies somewhere else. A number of groups worldwide are involved in a contest to find such explanation. But before looking for alternative explanations, it would be helpful to identify whether we can indeed speak about the two states of ions, what is the exchange/balance between them, and how they contribute to conductance of RTILs, as well as to the formation of the double layer.

The cluster concept in RTILs is actually not new. In the paper by Hu and Margulis (23) the idea of long living clusters of ions has been exploited for the interpretation of the so called "red-edge effect" (REE) observed in the study of fluorescence of the organic probe 2-amino-7-nitrofluorene dissolved in 1-butyl-3-methylimidazolium hexafluorophosphate ([Bmim][PF$_6$]). The focus of that work was on establishing the relation between REE and dynamic heterogeneity in RTILs, the latter shown to be crucial for the interpretation of the data. But in the course of that study, interesting features were revealed concerning the modes of ion transport. Analyzing the van Hove self-diffusion correlation function and comparing them with those predicted by classical Fickian (Gaussian) diffusion (24), they found that most ions diffuse much slower than expected from the Gaussian diffusion, but there are a group of ions that diffuse much faster. It was also concluded that there is a poor correlation between motions of the two kinds of ions, whereas in each group the correlations are strong.

Next, in some protic RTILs, using electrospray ionization mass spectrometry (ESI-MS), Kennedy and Drummond (25) observed the formation of aggregates of ions, the size of which depends on the nature of the cation and anion. They concluded that RTILs with strong tendency for ion clustering can be classified as "poor ionic liquids", in analogy with weak, poorly dissociating electrolytes. Generally, ESI-MS has been the principal technique employed to corroborate the ion cluster model, in which the bulk structure is depicted as a sea of polydisperse aggregates (26, 27). Thus, there is sufficient experimental evidence in favor of existence of clustering in RTILs but there is no unified view on the scale of this effect, which may vary from liquid to liquid.

Independent of data interpretation, the papers of Gebbie et al. had an immediate effect. The interpretation of compacity was reconsidered and the modified mean-field theory of electrical double layer developed (22, 28). Indeed, for given environmental condition (temperature, pressure, air humidity), $\gamma$ can be considered as a set parameter for each RTIL in the bulk. However, near the interface, across the electrical double layer, concentration of charge carriers may vary. Indeed, in the inhomogeneous electric field the balance from the clustered state to the free state of ions shifts in favor of the free state. Thus the distribution of potential and ions across the double layer must be considered self-consistently (22).

Recently Gavish et al. (29) and Rotenberg et al. (30) suggested explanations of the underscreening effect in concentrated electrolytes, based, respectively, on phenomenological density functional approach and mean-spherical approximation. At high electrolyte concentrations, these authors have demonstrated the emergence at the interface (29) and in the bulk (30), of self-assembled structures, manifested in extended oscillating internal domain in the spatial ion-ion correlation functions and the long monotonous tail beyond the oscillation range. But the tail did not come out as long as seen in experiments.

Generally, the crossover between decaying oscillatory profiles and monotonic decay of charge density correlation functions and screened potentials with increase of concentration of ions – known as Kirkwood crossover – has a long history. Kirkwood first predicted it, based on his analysis of the potential of mean-force in electrolytes (31). For the primitive model of electrolyte, using Ornstein-Zernike equation, Attard has studied the analytical laws for asymptotic behavior of the direct correlation function, as well as numerically its full behavior, using hypernetted chain approximation (32). He found, within the validity of approximations used, that (i) monotonically decaying screening length is slightly increased with respect to Debye length, and (ii) at high concentrations the oscillations with the period shorter than the monotonic decay length emerge. Earlier, Kjellander and Mitchell came to similar conclusions based on their charge renormalization analysis (33).

The effect has its background in the basic structure of correlations in dense ionic fluids, in which, with increase of ion concentration, ion pair spatial correlation functions change from monotonic exponential decay to decaying oscillations, formally as a result of a shift of the pole in their Fourier transforms from imaginary axis to a complex plane. This has been clearly demonstrated and analyzed within the generalized mean spherical approximation of a restricted primitive model of electrolyte by Leote de Carvalho and Evans (34). They have also studied phase diagrams with transition lines on the temperature-concentration plane, corresponding to this phenomenon (34). Kjellander (35) has just published a detailed, updated analysis of the emergence of renormalized and oscillating mode screening, discussing the "transient associations of each ion with several ions of opposite charge, which have high probability to be in the immediate neighborhood due to Coulomb attractions and effects of many-body correlations in the fluid". He speculates about the possibility of emergence of very long decay length caused by the effect of ion pairing. In another recent paper, Ciach (36) presented a density functional theory for ionic liquid or ionic liquid mixture with solvent molecules near a charged wall, in the linear response approximation. That theory shows the decaying oscillatory profiles of the potential distribution; however, the analytical expression for the envelope-decay-length does not give clear indications when these can get abnormally long.

Another way of looking at oscillating *vs* monotonic patterns of ionic correlation functions is to consider them as a manifestation of the clustered *vs* free ions dynamics assemblies (c.f. similar ideas of Kjellander (35)). Indeed, in simple terms, clusters determine the oscillating domains – they imply oscillations of charge density in them; free ions contribute to tails. Similarly, estimating the portion of ions contained in the former and in the latter could give independent evidence on the degree of clustering in RTILs. Somehow further studies should establish a detailed picture of ionic states in the bulk.

The concept of ion paring and clustering in conductivity measurements and simulations of RTILs has its own history. The conductivity roughly obeys the Nernst-Einstein relationship with the measured diffusion coefficient, but with some deviations that lead to smaller conductivities (37). The concept of ionicity suggested that this deviation is due to ion correlations and ion pairing. This interpretation has been subject to harsh scrutiny, with some authors suggesting that the concept of ion pairing and clustering in RTILs is not required to explain the deviation from the Nernst-Einstein (38, 39). On the other hand, some studies were in support of the concept of ion pairing and clustering in RTILs (40). Although there have been investigations into the properties of ion pairs in RTILs (41), there have been no reports, at least as far as the authors are aware, that estimate the fraction of ions in a state that contribute to conductivity. This is the gap we aim to bridge in the present study.

The main outstanding questions here are:

(i) What is the mechanism of ionic diffusion and conduction in RTILs?
(ii) What is the portion of ions that can 'freely' move and dominate the conductance and ionic screening?
(iii) Is there contribution to conductance of slowly diffusing ions corresponding to a 'bound' state?

(iv) Is the idealized picture of two states for ions physically justified for interpretation of experimental data?

(v) How easily does the exchange between the two states occur and what is the 'band gap' between them?

The present paper is focused on answering these questions. Our method of choice is molecular dynamics (MD) simulation. It allows to trace the trajectory of each individual ion, as well as to obtain the statistical characteristics of this motion such as self-diffusion coefficients, conductivity and alike, and to investigate their temperature dependence. While in the main text, we present the results only for one example RTIL, 1-n-butyl-3-methylimidazolium bis(trifluoromethanesulfonyl)imide ([Bmim][TFSI]), in the *Supporting Information* (SI) we show very similar results for different RTIL, as well as test the sensitivity of results to variation of the force-fields. The conclusions remain unchanged. Still, the answers that we will give cannot be considered final, but they set a framework for future experiments necessary for obtaining the ultimate answers.

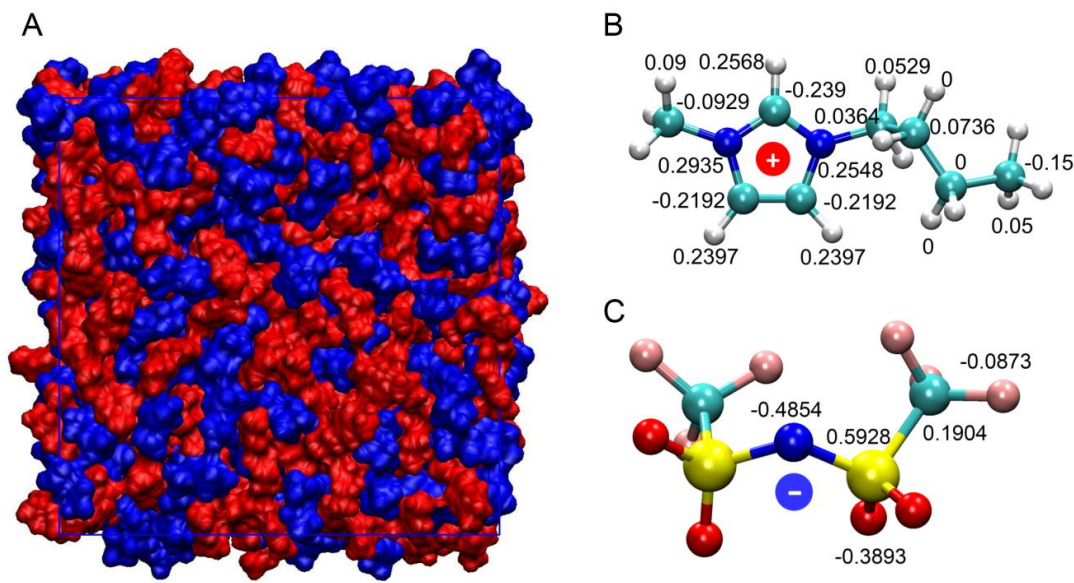

**Fig. 1. Molecular dynamics simulation of RTIL [Bmim][TFSI].** (A) Snapshot of MD simulation (red and blue colors indicate cations and anions, respectively). Molecular structure of (B) cation $Bmim^+$ and (C) anion $TFSI^-$. The numbers denote the partial charges on each atom in the all-atom model of ions used for MD modeling (unit: elementary charge). The solid red circle with the sign '+' represents the cation centre (i.e., the mass centre of the ring), and solid blue circle with the sign '-' indicates anion centre (i.e., the mass centre of the whole anion).

**Results and Analysis**

**Trajectory density approach**. MD simulations of RTIL [Bmim][TFSI] were performed at different temperatures in bulk system (Fig. 1, simulation details could be found in **Methods**). Respectively in Fig. 2A and 2B the trajectory of the centre of one cation and anion, which was arbitrarily chosen, is displayed. Considering diffusion of individual ions in a crowded environment can provide decisive information on the structure and dynamics of the entire system (42). It can be seen that an ion moves for a while re-entrantly (in anticorrelated fashion) in a confined volume and then 'speeds' up – undergoes correlated, persistent motion outside that volume to later get trapped in a new domain. Such picture can be viewed by denser trajectory clouds connected by sparse ones. The box counting method (43) was adopted to obtain the local trajectory density (i.e., the number of trajectory points in the cubicle divided by its volume, Fig. 2C-D), by dividing the simulation cell into elementary cubicles with a size of 0.3 nm (cf. Fig. S1). To differentiate the ion states, we conventionally assume the ion to be in a 'free state' when the local trajectory-density is smaller than the average density of all trajectories; otherwise the ion is considered to be in a 'bound state'. Therefore, the percentage of thus defined free ions, which if we neglect free voids it could equalize to $\gamma$, was computed based on the times ions spend in the free and in the bound state (having averaged the trajectories separately over all cations and all anions in the simulation). The percentage of free ions is shown to increase with temperature (Fig. 2E).

That temperature dependence is treated in terms of a Fermi-like distribution, an approximate expression for the fraction of free ions was derived in Ref. (28):

$$\gamma = \frac{1}{1+\frac{1}{2}e^{\frac{E_g}{k_BT}-\frac{\Delta S}{k_B}}} \tag{1}$$

where $E_g$ is the energy gap between the free and bound states, and $\Delta S$ is the entropy change when an ion transfers from bound to free state. Rewriting this equation as $\ln\left[\frac{1}{\gamma}-1\right]=-\left(\ln 2+\frac{\Delta S}{k_B}\right)+\frac{E_g}{k_BT}$, by plotting the simulated values of $\ln\left[\frac{1}{\gamma}-1\right]$ vs $\frac{1}{T}$, we get the gap energy from the slope and entropy from the intercept. That fitting is shown in Fig. 2F, giving 2.5 kJ/mol (0.026 eV) for cations and 2.6 kJ/mol (0.027 eV) for anions for the change in energy between the two states. The entropy change is small, roughly -0.012 kJ/(mol·K) for both cation and anion. Interestingly, it is negative, indicating that the ions have more degrees of freedom in the bound state than in the free state. That entropic change comprises relatively large contribution to the free energy of transfer: at 300 K it is 3.6 kJ/mol, and at 600 K is 7.2 kJ/mol for a cation. Correspondingly, this results in a free energy of transfer 6.1 kJ/mol (0.063 eV) and 9.7 kJ/mol (0.1 eV) at 300 K and 600 K, respectively. Within the accuracy of error bars, the intercept can vary within some 10%, which will affect these numbers rather than the trends. Similar phenomenon occurs in another RTIL [Bmim][$PF_6$] (Fig. S2).

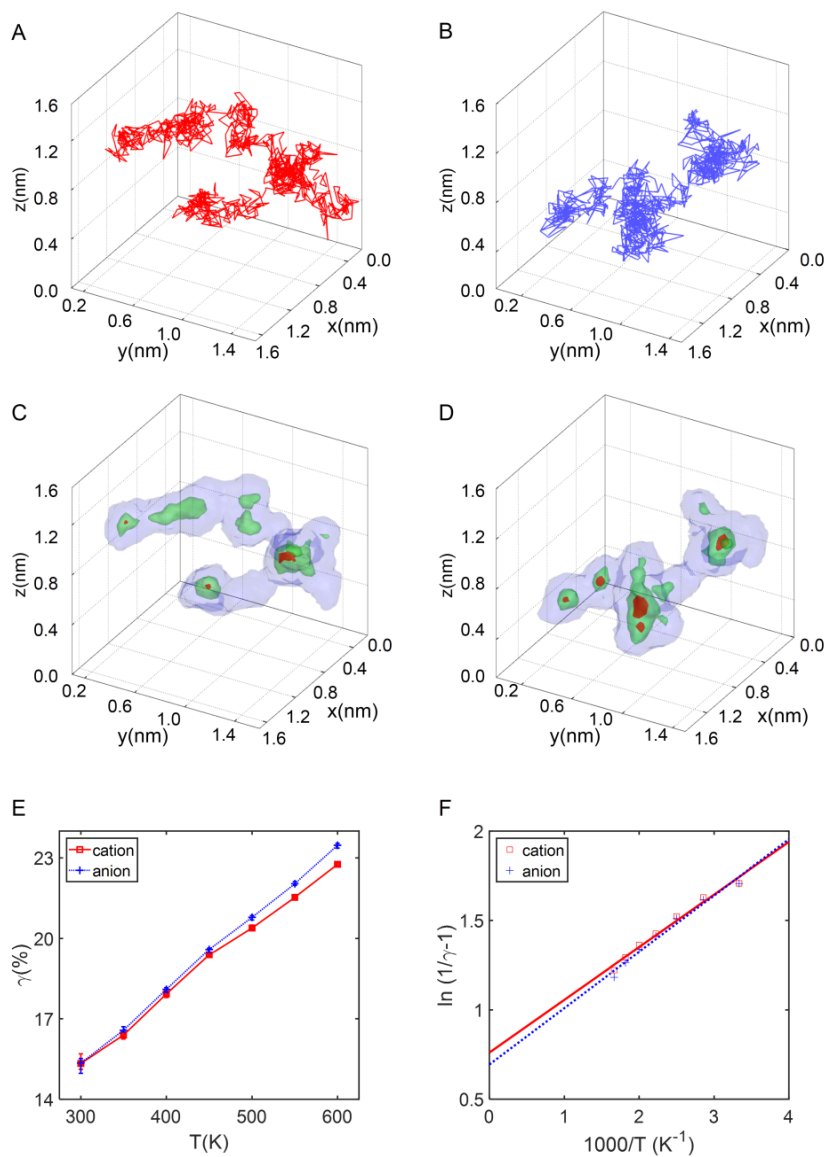

**Fig. 2. Two states of ion dynamics revealed by ion trajectories**, based on MD simulations of [Bmim][TFSI]. Trajectories of arbitrarily chosen (A) cation and (B) anion and the contour map of trajectory density of the (C) cation and (D) anion. (E) The percentage of free ion ($\gamma$-fraction given in percentage), as a function of temperature (lines are guide for the eye). (F) Fermi-like plot for the percentage of free ions, based on Eq. (1) (lines are the linear fit).

To characterize the ion dynamics in free and bound states, the **v**elocity **a**uto**c**orrelation **f**unctions (VACFs) were computed for free and bound ions in Fig. 3A, compared with those based on the whole trajectory analysis (i.e., when all states are taken together). VACF profiles show damped oscillations, in principle decaying to zero as expected, with the oscillations less pronounced as temperature increases. But for cations (Fig. 3A, *top panel*), VACFs exhibit an intermediate negative asymptotic plateau, which may be attributed to backscattering due to cation collisions and bouncing back and forth within an ion 'cage' in the bound state. With temperature increasing, the backscattering is weaker, probably ascribed to easier escape of cations from the cage. Similar trends hold for VACFs of free ions and bound ions, while bound ions exhibit more pronounced oscillatory motions than free ions. The diffusion coefficients of ions, D, in different dynamics states are obtained based on VACFs using the Green-Kubo integral formula (44), as shown in Fig. 3B (*top panel* for cations and *bottom panel* anions, respectively). The ions in free state diffuse much faster than in bound state and with temperature increasing, such deviation becomes more distinct for both cations and anions. Diffusion coefficient as a function of temperature were fitted by Arrhenius equation, resulting in an activation energy of ~21 kJ/mol (0.22 eV) for free ions, larger than that for bound ions, ~15 kJ/mol (0.16 eV) in Table 1. This is not surprising: individual ions in a clustered state do not move far and each elementary act of ion transfer requires less reorganization of the environment, as compared to long jumps of free ions. But the motions of ions in a clustered state are anticorrelated, indicating that the presence of confining potential makes diffusion less effective.

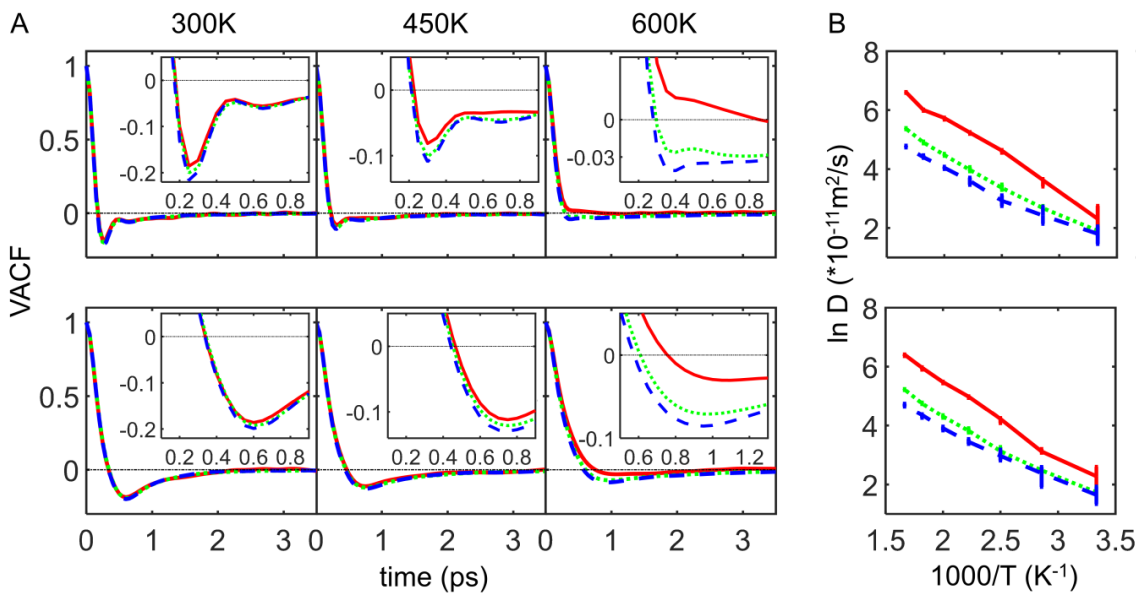

**Fig. 3. Temperature effects on diffusivity of ions**, studied by MD simulations of [Bmim][TFSI] (A) Velocity autocorrelation functions of cations (top panel) and anions (bottom panel), at indicated temperatures. (B) The diffusion coefficients of cations (top panel) and anions (bottom panel) as a function of temperature, shown in Arrhenius coordinates. Red-solid, green-dotted and blue-dashed lines represent free ions, overall ions, and bound ions, respectively.

**Table 1.** Arrhenius equation fitting of diffusion coefficients with different states.

$$D = D_0 \exp(-E/k_B T)$$

|  | free ions | | all ions | | bound ions | |
| --- | --- | --- | --- | --- | --- | --- |
|  | $D_0$ | $E$ | $D_0$ | $E$ | $D_0$ | $E$ |
| cation | 4.50 | 20.72 | 0.58 | 17.21 | 0.21 | 15.00 |
| anion | 3.85 | 21.03 | 0.51 | 17.39 | 0.20 | 15.05 |

Unit of $D_0$: $10^{-7} m^2/s$; unit of $E$: kJ/mol.

**Eulerian View on Ion Association** (Analysis based on ion pairing). The above concepts of 'free' and 'state' ions rise from the analysis of temporal trajectory, by tracing each ion moving. That approach did not specify how the ion is bound with each other. We may try to use also another approach based on conventional definition of ion pairing. Considering the fact that in the bulk RTILs ions (cations and anions) strongly interact with each other, the association among them, leading to partial neutralization in RTILs, could be described in terms of ion pairing (41, 45). This concept is clearly justified in electrolytic solutions; while in dense ionic systems, such as RTILs, it is hard to distinguish 'who is paired with whom', and whether pairs are not just parts of bigger clusters. Nevertheless, let us define the cation and anion as an ion pair when their ion centers are within a certain distance from each other, taking such distance to be the sum of the radii of the oppositely charged ions based on ring-anion spatial distribution function. Using free volume method (46), the radius of cation ring is computed as 0.275 nm (0.326 nm for anion), in line with the literature (47). Specifically, each ion can only form an ion pair with a counterion, behaving as a quasi-solvent molecule, while ions that are not involved in ion pairing would be considered as 'free'. With this approach [in analogy with a renowned analysis of the portion of hydrogen bonded molecules in water (48)], the percentage of free ions was computed. Remarkably, the results show similar trend as that quantified via trajectory analysis (Fig. S3).

In brief, approaches based on trajectory density and ion pairing analyses *both* unravel the same dynamic behavior of RTILs, that is, the free ions are in minority (within about 15% ~ 25%) of all ions in RTILs and their percentage increases with temperature.

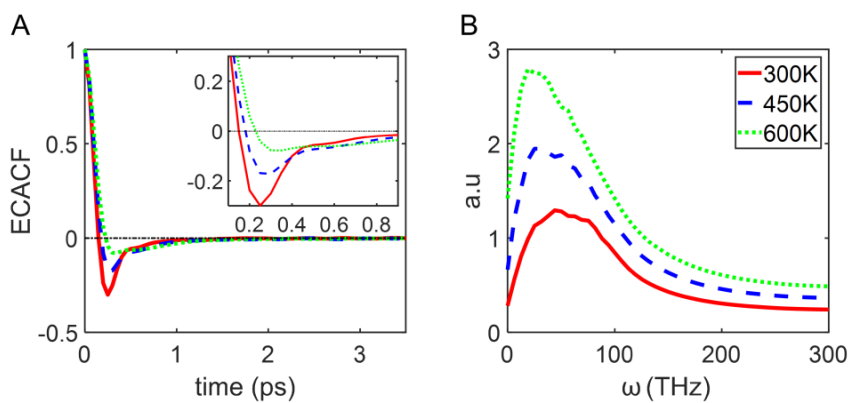

**Fig. 4. Temperature effects on (A) the electrical current autocorrelation function (ECACF) and (B) spectrum of frequency-dependent electrical conductivity**, both calculated from MD simulations of RTIL [Bmim][TFSI].

**Conductivity of RTILs.** The **e**lectric **c**urrent **a**uto**c**orrelation **f**unction (ECACF) was calculated to obtain the RTILs conductivity, since ECACF takes into account not only the ionic motions but also the contributions of cross-correlations among ions (45). It can be found that ECACFs decay very quickly to zero within 1 ps (Fig. 4A) and the conductivity spectrum shows a higher peak shifting toward low frequency as the temperature increases (Fig. 4B). Data for all studied temperatures can be seen in Fig. S4.

In view of the Green-Kubo integral of ECACF (44), including the contributions of ionic correlations, the electrical conductivity, $\Lambda$, is computed for the RTIL [Bmim][TFSI] and exhibits good agreement with experimental measurements (49), as shown in Fig. 5 (the red squares vs the blue dots). Several research groups reported that the electrical conductivity would be overestimated by Nernst-Einstein equation, because it neglects ionic association and assumes that each ion contributes one elementary charge (37, 44, 45, 50). Such overestimation would be clearly seen in Fig. 5, comparing the conductivity obtained by full ECACFs with that calculated from Nernst-Einstein equation (the red squares vs the black circles).

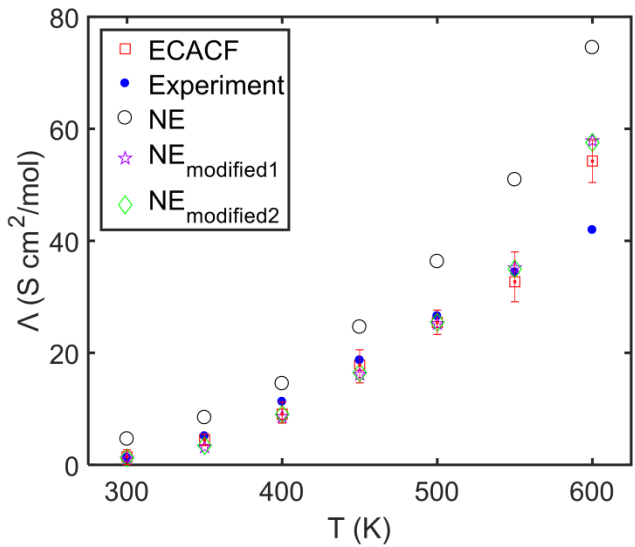

**Fig. 5. The electrical conductivity, $\Lambda$, of [Bmim][TFSI] as a function of temperature**. The red square shows the values calculated from the MD-simulated electric current autocorrelation function (Fig. 4A). The blue dots represent experimental data from Ref. (49). The black circles display the values computed via original Nernst-Einstein equation with diffusion coefficients shown in Figure 3B. The magenta stars are based on Eq. (2), with free ion percentage evaluated by the trajectory density method. The green diamonds are based on Eq. (2), with free ion percentage evaluated by ion pairing method.

With the above analysis of free and bound ions, let us assume that the free ions dominantly contribute electrical conductivity, and the Nernst-Einstein equation could be modified as,

$$\Lambda = \frac{1}{k_BT}\sum_i \alpha_i q_i^2 D_i \quad (2)$$

where $\alpha_i$, $q_i$ and $D_i$ represent the portion of free ions, the charge, and diffusion coefficient of ion species i (cation and anion). With the help of this modified Nernst-Einstein equation, the electrical conductivity was computed with the introduced parameter of $\alpha_i$ (see purple stars in Fig. 5), which shows remarkable agreement with the approach of the Green-Kubo integral based on full ECACFs as well as the experimental data (49). This modified Nernst-Einstein equation works well also for another RTIL [Bmim][PF$_6$] (Fig. S5). Similarly, using in Eq. (2) the portion of free ions obtained from the ion pairing method (see green diamonds in Fig. 5) gives results close to those of the exact calculation.

Why is this so? Neglecting ionic correlations between free ions, intrinsic into the Nernst-Einstein equation, seemingly overestimates the electrical conductivity. On the other hand, ignoring the contribution from bound ions would underestimate the conductivity. Mutual compensation of these two effects is presumably responsible for coincidence of the data obtained from the calculation of the full ECACFs and the modified Nernst-Einstein equation. Therefore, it can be concluded that the free ions dominate the electrical conductivity of bulk RTILs.

Note that the values of the conductivity obtained from our simulation are surprisingly close to the experimentally measured ones for RTIL [Bmim][TFSI] (49) and, as shown in Fig. S5, also for RTIL [Bmim][PF$_6$] (51).

**Discussion and Conclusions**

**Mechanisms of Ion Transport in RTILs**. There were different ideas about the mechanisms of ion transport in RTILs. One of the scenarios was proposed by Abbot (52, 53) who assumed that in such a concentrated ionic systems, only a hole mechanism of transport would be feasible. The findings presented above cannot directly prove or disprove that hypothesis, as looking at any individual trajectory, we cannot tell how each elementary step of ion transfer proceeds. Another, often discussed, such scenario is a quasi-Grotthuss (relay) mechanism, in which the ions in each elementary act do not move far, but each ion just slightly shifting, kicking another ion to continue. Whereas this is not excluded for ions inside the clusters, our free ions can seemingly individually move considerable distances, as we can see it from their individual trajectories.

**Ionic Semiconductor Analogy**. As typical in condensed matter physics, in complex cases people indulge into the language of 'collective' phenomena. It is, thus, tempting to treat RTILs as an 'ionic semiconductor', in which in the ground state ions populate the ionic 'valence' (or excitonic') band, but they can be excited into a 'conductance' band (54). However, the band gap energy that we obtained in the fits of Fig. 2F is very small, $E_g =$ 0.026 eV (which is approximately 1 $k_B T$ at room temperature) but there is a compatible positive contribution to the free energy of transfer coming from the negative entropy change.

Ionic semiconductor analogy for RTILs was raised in Ref. (7) and in the context of discussion of the results of Ref. (13); our findings show that the such concept makes sense as simple approximation to a complex problem. A similar idea was discussed in 1970s-80s in interpretation of ion transport in superionic conductors' single charge carrier solid electrolytes (55, 56). For instance, in $Ag_4RbI_5$ the conducting ions are silver cations, the rest of the ions forming a solid lattice in which $Ag^+$ ions move. But the latter were speculated to occupy two states: ground, immobile states in the crystal unit cells, and an excited state in a conduction band in which they move collectively. The band gap between them determines the activation energy of conductance that has a clear Arrhenius dependence in a wide range of temperatures with small activation energy of 0.069 eV (57). Of course, RTILs are much more complex systems than crystalline superionics: there is no lattice in RTILs, both ions are capable to move; but the picture of long trajectories in the motion of free ions does not contradict to the ionic semi-conductor concept. Although we trace the path of an individual ion, its trajectory is a part of a cooperative motion of the ions, in free and bound states, and reflects collective dynamics of entire ionic liquid.

More experiments are needed to verify this concept. Measurements of the Hall effect come to mind first. However, the latter is difficult from experimental point of view (58, 59), as well as is not straightforward in interpretation (60, 61).

Our last note is on possible implications for the double layer theory. If the exchange between the conduction and valence bands is much faster than the RC-time of charging the double layers, which is practically always the case, it would be legitimate to speak about the average concentration of mobile charge carriers that contribute to the formation of the electrical double layer. But if even only 20% of ions were on average free and giving the major contribution to the double layer, it is already a very concentrated ionic system, and the spatial structure of the double layer, if described in all its complexity, will presumably contain oscillating ('overscreening') part and the exponential tail ('underscreening' part). In fact, a recent study of the temperature dependence of the differential capacitance found the shape of the capacitance curve to change as if ion pairs and clusters were breaking up with increasing temperature (28), which is in accordance with the qualitative results here (although the reported values of $\gamma$ therein were slightly different). Bound ions will contribute not only to effective dielectric constant of the system, but also to the spatial potential and charge distributions.

**Methods**

As shown in Fig. 1, MD simulation of RTIL [Bmim][TFSI] was performed using a customized MD code Gromacs (62). The force fields of [Bmim][TFSI] were adopted from Ref. (63). A three-dimensional, periodic cubic simulation system consisting 300 ion pairs was simulated, with the temperature maintained in the interval of 300-600 K (in steps of 50K) using the Nose-Hoover thermostat with a relaxation time of 0.2 ps. The Parrinello-Rahman barostat was taken to maintain the pressure of 1 bar with a relaxation time of 2 ps in NPT ensemble. The non-bonded interactions with repulsion and dispersion were computed using Buckingham potential. The electrostatic interactions were computed using the particle-mesh Ewald method. To compute the interactions in the reciprocal space, a fast Fourier transformation grid spacing of 0.12 nm and cubic interpolation for charge distribution ware used. A cutoff distance of 1.2 nm was used in the calculation of electrostatic interactions in the real space. The simulation box was initially equilibrated for 10 ns in the isothermal-isobaric (NPT) followed by canonical ensemble (NVT) for a 30 ns production run. To ensure the accuracy of the simulation results, each case was repeated three times with different initial configurations.

In addition, MD simulations containing 600 ion pairs of RTIL [Bmim][TFSI] were performed to test whether the simulation box size would affect the results. As shown in Fig. S6, it can be found that the simulation box size have ignorable influence on the results. Moreover,to show how conclusions obtained from the modeling of RTIL [Bmim][TFSI] are universal, another RTIL [Bmim][$PF_6$] was investigated with the same MD simulation setup, but different force fields adopted from Ref. (40).

ACKNOWLEDGMENTS

We are thankful to Gleb Oshanin, Susan Perkin and Tom Welton for useful discussions. GF, MC, and SB acknowledge the funding support from the National Natural Science Foundation of China (51406060) and Shenzhen Basic Research Project (JCYJ20170307171511292); AAK acknowledges the Leverhulme Trust for funding (RPG-2016-223). MU acknowledges the financial support of the Deutsche Forschungsgemeinschaft (DFG), grant no. BA 1008/21-1. ZG was supported through a studentship in the Centre for Doctoral Training on Theory and Simulation of Materials at Imperial College London funded by the EPSRC (EP/L015579/1) and from the Thomas Young Centre under grant number TYC-101.

# Supplementary Information

Fig. S1 shows the size distribution of different clusters of ion trajectory cloud (Fig. 2A-B) in bound state. One can see that the most trajectory cloud size is around 0.3 nm. Therefore, we adopted 0.3 nm as the optimal cubicle size for the box counting method.

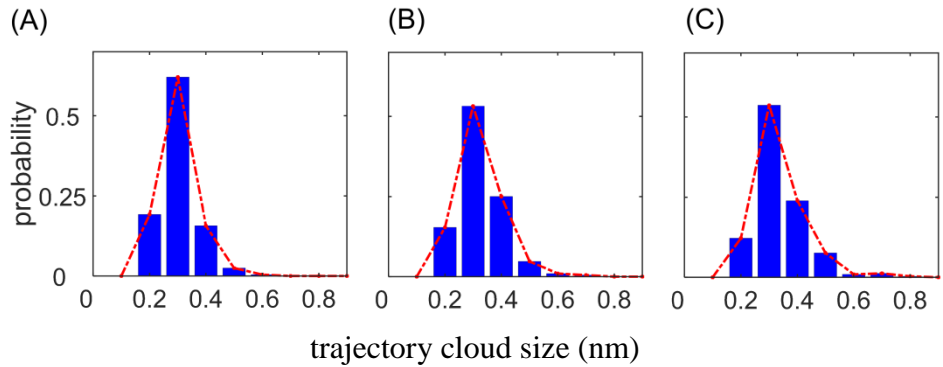

**Fig. S1. The effect of the cubicle size in box counting method on trajectory cloud size for bound ions.** Cubicle size is chosen as (A) 0.2 nm, (B) 0.3 nm, and (C) 0.4 nm.

In Fermi-like plot for the portion of free ions in [Bmim][PF$_6$] (Fig. S2B), the fitting, using Eq. (1), gives for the change in energy between the two states ('band gap') $E_g$ =2.86 kJ/mol (0.030 eV) for cations, and 2.62 kJ/mol (0.027 eV) for anions in RTIL. The entropy change is small, roughly -0.011 kJ/(mol·K) for both cation and anion.

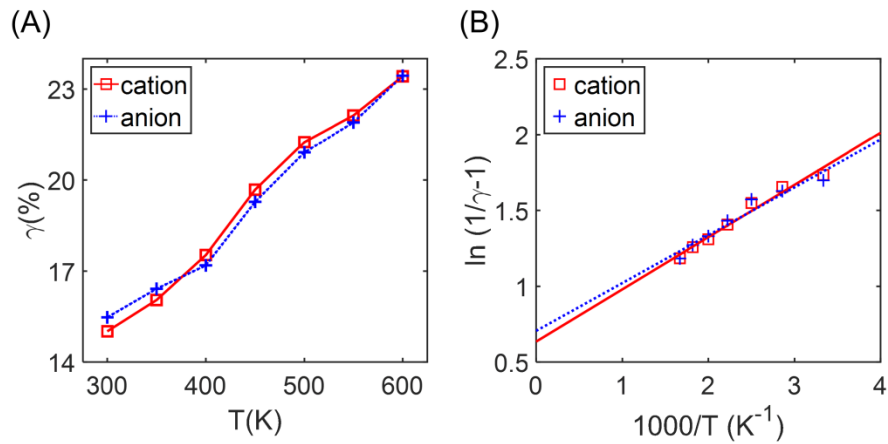

**Fig. S2. Free ions RTIL [Bmim][PF$_6$] as a function of temperature.** (A) The percentage of free ion ($\gamma$-fraction given in percentage), as a function of temperature (lines are guide for the eye). (B) Fermi-like plot for the portion of free ions, based on Eq. (1) (lines are the linear fit).

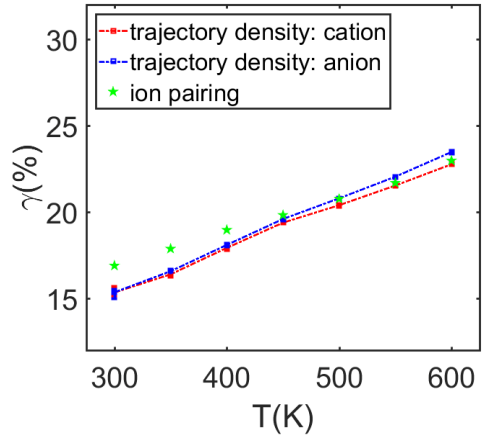

**Fig. S3. The comparison of free ion percentage based on trajectory density and ion pairing for RTIL [Bmim][TFSI].** The red square (cation) and blue square (anion) are the free ion based on trajectory density, and the green star is the free ion based on ion pairing.

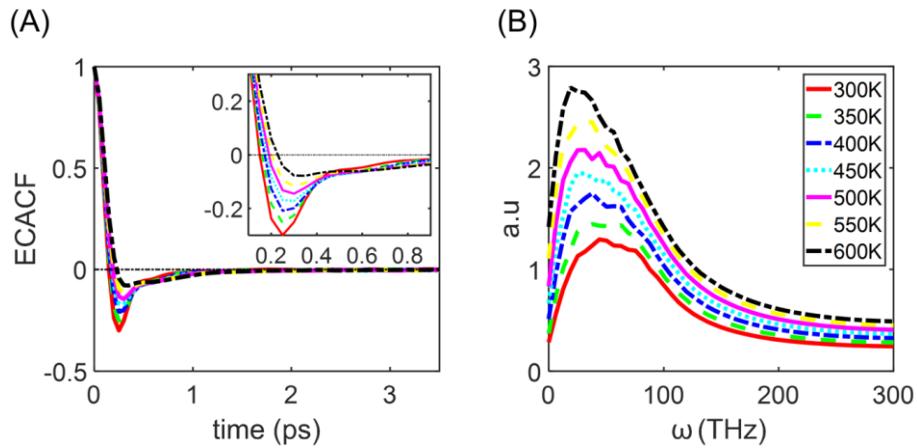

**Fig. S4. The electric current autocorrelation function (A) and conductivity spectrum (B) as a function of temperature,** both calculated from MD simulations of RTIL [Bmim][TFSI].

As Fig. S5 shows, modified Nernst-Einstein equation, taking into account the introduced parameter of free ion percentage (see Fig. S2A), gives the electrical conductivity, in line with the approach of the Green-Kubo integral based on full ECACFs as well as the experimental data (1).

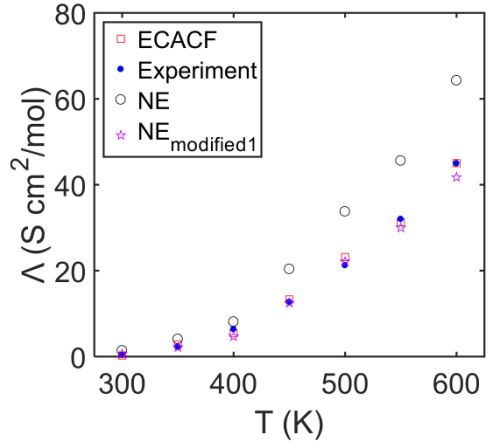

**Fig. S5. The electrical conductivity, Λ, of [Bmim][PF$_6$] as a function of temperature**. The red square show the values calculated from the MD-simulated electric current auto-correlation function. The blue dots represent experimental data from Ref. (1). The black circles display the values computed via original Nernst-Einstein equation. The magenta stars are based on Eq. (2), with free ion percentage evaluated by the trajectory density method.

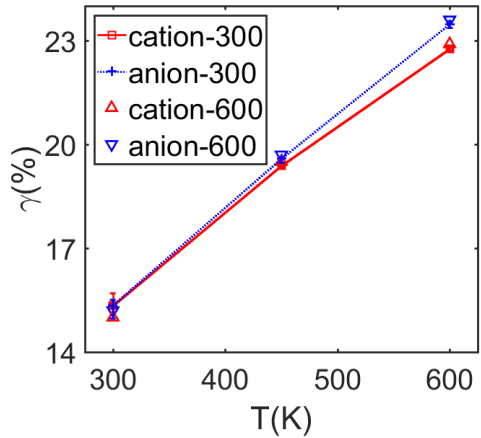

**Fig. S6. The comparison of free ion percentage for RTIL [Bmim][TFSI] modeled in different-sized MD simulation.** The red square (cation) and blue cross (anion) are obtained by MD simulation with 300 ion pairs of [Bmim][TFSI]; the red triangular (cation) and blue triangular (anion) are obtained by MD simulation with 600 ion pairs of [Bmim][TFSI].